\documentclass{epl}

\usepackage{amssymb}

\newcommand{\be}{\begin{equation}}
\newcommand{\ee}{\end{equation}}
\newcommand{\bea}{\begin{eqnarray}}
\newcommand{\eea}{\end{eqnarray}}
\newcommand{\ket}[1]{\ensuremath{\left| #1 \right\rangle}}
\newcommand{\bra}[1]{\ensuremath{\left\langle #1 \right|}}
\newcommand{\bracket}[3]{\ensuremath{\left\langle #1 \left| #2 \right| #3
    \right\rangle}}

\title{Tunable coupling of qubits: nonadiabatic corrections}
\shorttitle{Tunable coupling ...}
\author{Carsten Hutter\inst{1}, Alexander Shnirman\inst{1},
Yuriy Makhlin\inst{2}, and Gerd {Sch\"on}\inst{1}}
\institute{
\inst{1} Institut f\"{u}r Theoretische Festk\"{o}rperphysik and DFG-Center
    for Functional Nanostructures (CFN), Universit\"{a}t Karlsruhe,
    D-76128 Karlsruhe, Germany.\\
\inst{2} Landau Institute for Theoretical Physics, Kosygin st. 2, 119334
    Moscow, Russia.}


\begin{document}

\maketitle

\begin{abstract}
We analyze the coupling of qubits mediated by a tunable and fast element beyond the adiabatic approximation.
The nonadiabatic corrections are important and even dominant in parts of the relevant parameter range. As an
example, we consider the tunable capacitive coupling between two charge qubits mediated by a gated Josephson
junction, as suggested by Averin and Bruder.
The nonadiabatic, inductive contribution persists when the capacitive coupling is tuned to zero. On the other
hand, the total coupling can be turned off (in the rotating wave approximation) if the qubits are operated
at symmetry points.
\end{abstract}

\section{Introduction}

Most approaches to quantum computation rely on tunable interactions between pairs of qubits. These should be
turned off for independent single-qubit manipulations and to reduce crosstalk of qubits during readout, and
they should be turned on selectively in a controlled way in order to perform two-qubit logic gates. In
principle, refocussing techniques, known from NMR, can be used to suppress the effect of fixed (uncontrolled)
couplings, but they require considerable overhead and precision of the pulses. Thus tunable setups, which
allow minimizing the residual couplings, are desired.
At this stage, most experiments with Josephson qubits~\cite{nature2003pashkin,science2003berkley,%
nature2003yamamoto,prl2004izmalkov,prl2005majer,science2005mcdermott}
have been performed with direct and fixed couplings. But various schemes with tunable couplings have been
proposed~\cite{Our_Nature,You_commuting,%
Plastina_Falci,Blais_Tunable,Averin_Bruder,ieee2003filippov,Lantz,Plourde,%
Blais_CQED,Maassen_Mediated,Bertet_Parametric,Nakamura_Tunable,prl2005rigetti,prl2006liu}. Some of them gain
their tunability from ac-driving \cite{prl2005rigetti,prl2006liu}, but the majority relies on additional
circuit elements, such as switchable Josephson junctions, inductors, LC circuits (cavities), or further qubits.
One can distinguish two operation principles: For ``resonant''
couplers~\cite{Plastina_Falci,Blais_Tunable,Blais_CQED} the coupling element, typically an oscillator, is
tuned into resonance with one or both qubits. Alternatively one can use ``adiabatic''
couplers~\cite{Our_Nature,You_commuting,Averin_Bruder,Lantz,Plourde,%
Maassen_Mediated,Bertet_Parametric,Nakamura_Tunable,ieee2003filippov},
where the coupling element has a much higher excitation energy than the qubits and remains in its ground state
while mediating the coupling. In the following we will concentrate on these adiabatic coupling schemes.

Some early proposals \cite{Our_Nature,You_commuting} made use of a fixed coupler, but gained tunability by
using SQUID-type qubits with flux-controlled Josephson energies. They suffer from the difficulty that the
coupling can be switched off only if one succeeds in fabricating identical junctions.
The alternative approaches, proposed more recently, employ a tunable coupler which -- ideally -- can be tuned to
cancel fixed existing couplings~\cite{Averin_Bruder,Plourde,Nakamura_Tunable}.
In addition, by modulating the coupling constant around such ``zero'' points one can tune two qubits with
different energy splittings into resonance, while keeping both at their symmetry points where decoherence
effects are minimized~\cite{Bertet_Parametric,Nakamura_Tunable}.
In this Letter we will show that nonadiabatic corrections are important around such ``zero'' coupling points.
For example, we find that a gated Josephson junction produces, in addition to a tunable capacitive coupling,
an inductive one, which dominates when the capacitive interaction is switched off. The importance of inductive
corrections in the charging regime of Josephson junctions has recently been pointed out also by
Zorin~\cite{Zorin_InductiveJJ}. Certain nonadiabatic corrections were also noted in
ref.~\cite{Nakamura_Tunable}.

Below, we first present a general theory of tunable adiabatic coupling. As specific example we analyze the
setup with tunable capacitive coupling proposed by Averin and Bruder~\cite{Averin_Bruder} and demonstrate the
importance of inductive corrections in different regimes. When the charging energy of the coupling junction
dominates over the Josephson energy, $E_C\gg E_{\rm J}$, the inductive coupling is weak, but is important when
the capacitive interaction is tuned to zero. In the opposite limit, $E_{\rm J}\gg E_C$, the inductive coupling
always dominates over the weak capacitive interaction, and we recover earlier results for inductively coupled
charge qubits~\cite{Our_PRL}. In the present discussion we concentrate on charge qubits, but our general
formulation and conclusions apply equally to other types of qubits and couplers.

\section{General theory}

We consider a system of two qubits coupled via an adiabatic coupler and described by the Hamiltonian
\be
  \label{eq:Hfull}
  H=H_0+V=H_{\rm qubits} + H_{\rm coupler}(\gamma) + V \ .
\ee
The coupler is controlled by a parameter $\gamma$. We assume an interaction of the form $V = \lambda A B$,
where $\lambda$ is the coupling constant, $A$ an observable of the coupler, and $B$ an arbitrary function
of observables of  both qubits. In general $H_{\rm qubits}$ includes a direct and fixed coupling between qubits.
In the following we consider the situation, where this fixed coupling is (nearly) canceled by tuning the
coupler appropriately.

The coupler Hamiltonian can be brought into diagonal form,
$H_{\rm coupler} = \sum_{m=0}^{\infty} E_m \ket{m}\bra{m}$, where both the eigenstates $\ket{m}$ and energies
$E_m$ depend on $\gamma$. The small parameter governing the adiabatic approximation is
$|H_{\rm qubits}|/(E_1 - E_0)$, where  $|H_{\rm qubits}|$ denotes the maximum energy difference between the
eigenstates of $H_{\rm qubits}$. To proceed we integrate out the coupler and derive an effective Hamiltonian
in the subspace of slow degrees of freedom of the qubits. The effective interaction $V_{\rm eff}$ is found by
projecting the full time-evolution operator
$S(t,0)=T\,\exp{[}- {\rm i} \int_{0}^{t} V_{\rm I}(t')dt'/\hbar{]}$ (in the interaction representation) onto
the ground state of the coupler, $S_{\rm eff}(t,0)\equiv \bra{0}S(t,0)\ket{0}$, and defining
\bea
  S_{\rm eff}(t,0) =
    T\,\exp{[}-\frac{{\rm i}}{\hbar}\int_{0}^{t}V_{\rm eff,I}(t')dt'{]}\ .
\eea
We assume the interaction to be weak, such that  $\lambda \langle A\rangle \langle B\rangle/(E_1-E_0)\ll 1$,
for all matrix elements of $A$ and $B$. Expanding the evolution operator $S_{\rm eff}(t,0)$ up to second order
we find
\bea
  \label{equ:S_eff}
  S_{\rm eff}\approx
  1&-&\frac{{\rm i} \lambda}{\hbar}\int_0^t
    \bracket{0}{A_{\rm I}(t')}{0} B_{\rm I}(t')dt'\nonumber\\
  &+&\left(\frac{{\rm i} \lambda}{\hbar}\right)^2\,
    \int_0^t\int_0^{t_1}
    \bracket{0}{A_{\rm I}(t_1)A_{\rm I}(t_2)}{0}B_{\rm I}(t_1)
    B_{\rm I}(t_2)\,dt_1\,dt_2\ .
\eea
The first order contribution to $V_{\rm eff,I}$ is thus given (in the Schr\"odinger representation) by
\be
  \label{eq:V1}
  V_{\rm eff}^{(1)} = \lambda \bracket{0}{A}{0} B\ .
\ee
To find the second order contribution we insert the unity, $\sum_m \ket{m}\bra{m}$, in the last term of
eq.~(\ref{equ:S_eff}) between the two coupler operators. The part with $m=0$ is the second order, reducible
term in the expansion of $\, T\,\exp{[}- {\rm i} \int_{0}^{t} V_{\rm eff,I}^{(1)}(t')dt'/\hbar{]}$.
The terms with $m \neq 0$ give rise to a new contribution $\tilde{V}_{\rm eff,I}^{(2)}$ to the effective
interaction Hamiltonian, defined by
\be
  \label{eq:mneq0terms}
  \int_0^t \tilde{V}_{\rm eff,I}^{(2)}(t')\, dt'=-i\frac{\lambda^2}{\hbar} \sum_{m=1}^\infty
  \left|\bracket{0}{A}{m}\right|^2 \cdot I_m \ ,
\ee
where
$
    I_m \equiv \int_0^t\int_0^{t_1}
    B_{\rm I}(t_1)B_{\rm I}(t_2)\,{\rm e}^{{\rm i}(E_0-E_m)(t_1-t_2)/\hbar}
    \,dt_1\,dt_2
$.
Introducing $\tau=t_1-t_2$ and $T=(t_1+t_2)/2$ we rewrite the integral as
\be
    I_m=\int_0^t dT \int_0^{F(T)} d\tau\,
    B_{\rm I}(T+\tau/2) B_{\rm I}(T-\tau/2)\,{\rm e}^{-{\rm i}\omega_{m0 }
    \tau}\ ,
\ee
where $\omega_{m0}=(E_m-E_0)/\hbar$ and $F(T)=2\, \min(T,t-T)$. Because of the assumed separation of time scales,
$B_{\rm I}$ varies slowly on the time scale $\omega_{m0}^{-1}$, and  can be expanded,
\bea
  I_m &\approx&
    \int_0^t dT \int_0^{F(T)} d\tau\,{\rm e}^{-{\rm i}\omega_{m0 } \tau}
    \nonumber\\
  &&\cdot\left[ B_{\rm I}(T)
      +\frac{\tau}{2} \dot{B}_{\rm I}(T) +\frac{\tau^2}{8}
      \ddot{B}_{\rm I}(T)+...\right]    \,
    \left[ B_{\rm I}(T)-\frac{\tau}{2} \dot{B}_{\rm I}(T)
      +\frac{\tau^2}{8} \ddot{B}_{\rm I}(T)+...\right] \ .
\eea
For $T \gg \omega_{m0}^{-1}$ we can set $F(T)\rightarrow \infty$. Substituting the resulting $I_m$ into
eq.~(\ref{eq:mneq0terms}) we find the second-order contribution to the effective Hamiltonian (in the
Schr\"odinger picture)
\bea
  \tilde{V}_{\rm eff}^{(2)} =
  -\frac{\lambda^2}{\hbar} \sum_{m=1}^\infty
  \left|\bracket{0}{A}{m}\right|^2
  \left[
    \frac{B^2}{\omega_{m0}}+\frac{{\rm i}[B,\dot{B}]}{2\omega_{m0}^2}
    +\frac{\left[2\dot{B}^2-\{B,\ddot{B}\}\right]}{4\omega_{m0}^3}+...
  \right]\ .
\eea
Note that the time derivatives in the Schr\"odinger picture should be understood as commutators, {\it e.g.},
$\dot B={\rm i} [H_{\rm qubits},B]/\hbar$.

The effective Hamiltonian so far is
$\tilde{H}_{\rm eff} \approx H_{\rm qubits} + V_{\rm eff}^{(1)} +  \tilde{V}_{\rm eff}^{(2)}$.
To simplify the last term of $\tilde{V}_{\rm eff}^{(2)}$ an ``integration by parts'' can be performed which
adds a full time derivative to $\tilde{V}_{\rm eff}^{(2)}$. This is achieved by a unitary transformation
$H_{\rm eff} = U \tilde{H}_{\rm eff} U^{-1}$, where $U=\exp{[}iQ{]}$ and
$
  Q = \sum_{m \ge 1} \frac{\lambda^2}{4\hbar^2\omega_{m0}^3}
    \left|\bracket{0}{A}{m}\right|^2\,\{B,\dot B\}\ 
$.
This yields $ V_{\rm eff}^{(2)} = \tilde V_{\rm eff}^{(2)} + i[Q,H_{\rm qubits}]$, {\it i.e.}
\bea
  \label{eq:V2}
   V_{\rm eff}^{(2)} =
  -\frac{\lambda^2}{\hbar} \sum_{m=1}^\infty \left|\bracket{0}{A}{m}\right|^2
  \cdot\left[
    \frac{B^2}{\omega_{m0}}+\frac{{\rm i}[B,\dot{B}]}{2\omega_{m0}^2}
    +\frac{\dot{B}^2}{\omega_{m0}^3}+...
  \right]\ .
\eea

The full effective Hamiltonian thus reads
\be
  \label{eq:Heff}
  {H}_{\rm eff} = H_{\rm qubits} + V_{\rm eff}^{(1)} +  V_{\rm eff}^{(2)}+O(\lambda^3)\ .
\ee
Together with eqs.~(\ref{eq:V1}) and (\ref{eq:V2}) it constitutes our main result.
The tunable part of the coupling, eq.~(\ref{eq:V2}), depends via $\ket{m(\gamma)}$ and $\omega_{m0}(\gamma)$
on the parameter $\gamma$. For time-independent qubit operators $B_I$ we recover the results of the
Born-Oppenheimer approximation. Indeed, eqs.~(\ref{eq:V1}) and (\ref{eq:V2}) (with $\dot B=0$) can be obtained
by expanding the ground-state energy $E_0(B)$ of the coupler and the interaction term $\lambda A B$ in
$\lambda$. In general $B_I(T)$ depends on time, and we obtain nonadiabatic corrections. In the considered
limit, they are smaller than the adiabatic contributions, however, they still dominate the {\it total} qubit
coupling, if the tunable adiabatic part in eq.~(\ref{eq:V2}) (approximately) cancels further direct coupling
terms in $H_{\rm qubits}$. Below we analyze the nonadiabatic corrections for specific systems.

\section{Tunable capacitive coupling}

\begin{figure}
\onefigure[width=8cm]{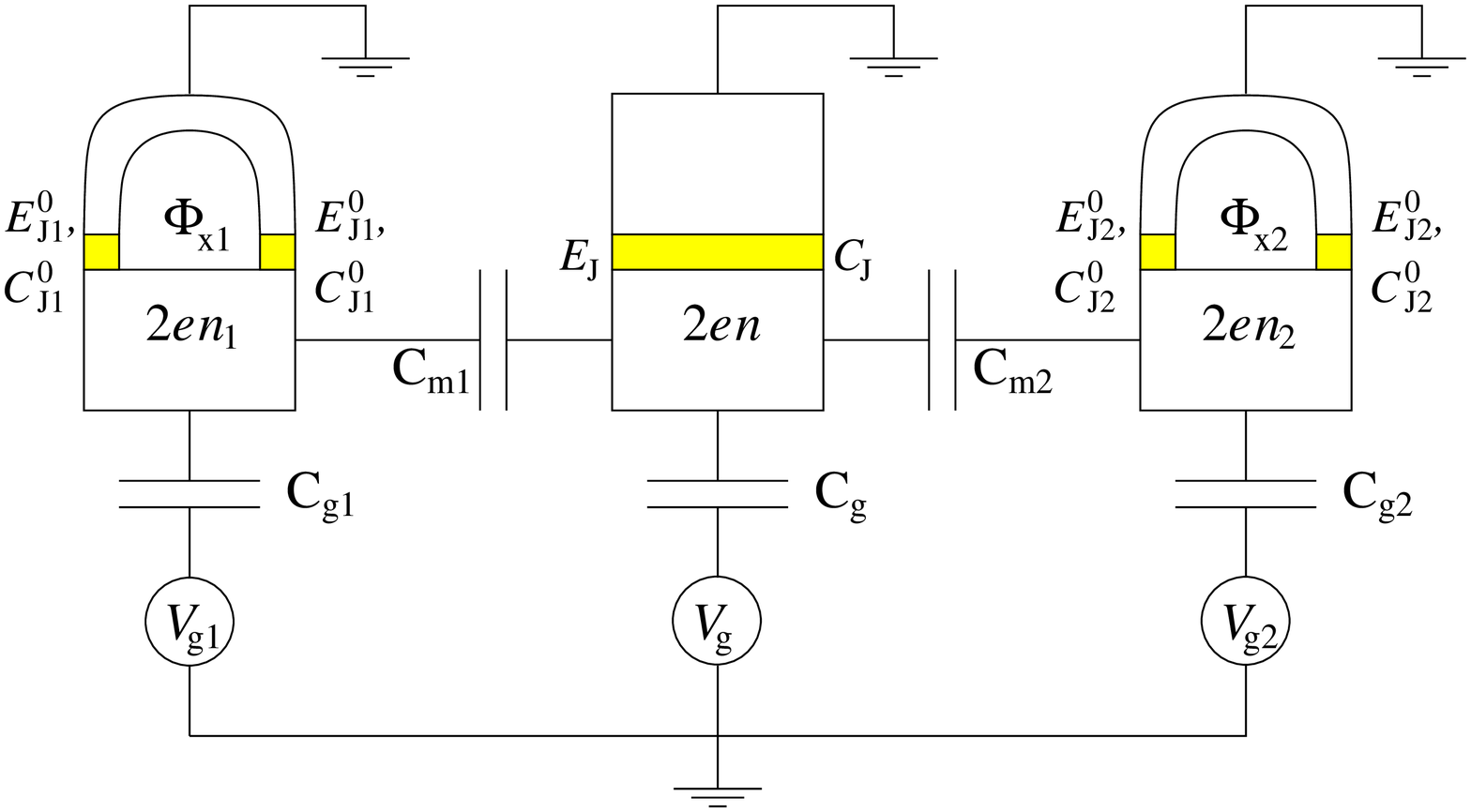}
\caption{Tunable capacitive coupler proposed in ref.~\cite{Averin_Bruder}. We slightly  modified the system by
  introducing the gate capacitor at the coupler. This influences the expressions for the gate charges but does
  not change the form of the Hamiltonian.
  The effective qubit Josephson energies can be tuned using control fluxes in the SQUID loops.
  To simplify expressions, we assume the SQUIDs to be symmetric.
  Then
  $E_{{\rm J}k} = 2 E_{{\rm J}k}^0\cos(\pi\Phi_{{\rm x}k}/\Phi_0)$ and
  $C_{{\rm J}k} = 2 C_{{\rm J}k}^0$, where $E_{{\rm J}k}^0$ is the Josephson
  energy of one of the SQUID's junctions and $C_{{\rm J}k}^0$ is its capacitance.
  The charging energies are given by
  $E_C=2e^2/(C_\Sigma-\sum_k C_{{\rm m}k}^2/C_{\Sigma k})$ and
  $E_{Ck}=2e^2/C_{\Sigma k}$, where
  $C_\Sigma = C_{\rm J} + C_{\rm g} + C_{{\rm m}1} + C_{{\rm m}2}$ and
  $C_{\Sigma k} = C_{{\rm J}k} + C_{{\rm g}k} + C_{{\rm m}k}$.}
\label{Figure:AverinBruder}
\end{figure}

The system shown in fig.~\ref{Figure:AverinBruder}, which is similar to the one proposed in
ref.~\cite{Averin_Bruder}, provides an example of tunable capacitive coupling. It is described by the
Hamiltonian
$
    H=E_C[n-n_{\rm g}-q(n_1,n_2)]^2-E_{\rm J}\cos\phi+\sum_{k=1,2} H_k
$
with single-qubit Hamiltonians $H_k=E_{Ck}(n_k-n_{{\rm g}k})^2-E_{{\rm J}k}\cos\phi_k $. Here $\phi$ and
$\phi_k$ are the phase differences across the coupling junction and of the qubits, respectively.
The (dimensionless) gate charges of the qubits only depend on the applied gate voltages,
$n_{{\rm g}k} = C_{{\rm g}k} V_{{\rm g}k}/2e$.
But the gate charge of the middle junction consists of $n_{\rm g}=C_{\rm g} V_{\rm g} / 2e$ plus the gate
charge induced by the two qubits,
$q(n_1,n_2)= -\sum_k (C_{{\rm m}k}/C_{\Sigma k})(n_k - n_{{\rm g}k})$.

In order to connect to  eq.~(\ref{eq:Hfull}) we rewrite the Hamiltonian of the system in the form
\bea
  H_{\rm qubits} &=& \sum_{k=1,2}H_k+E_C q^2+2E_C n_{\rm g} q\ , \nonumber\\
  H_{\rm coupler} &=& E_C (n-n_{\rm g})^2-E_{\rm J}\cos\phi\ , \nonumber\\
  V &=& -2E_C \, n \, q \ .
\eea
Thus we have $A=n$, $B=q(n_1,n_2)$, $\gamma=n_{\rm g}$, and $\lambda=-2E_C$, and the small parameter of the
perturbation theory is $c_k E_C/(\hbar\omega_{m0})$ with the small constants
$c_k \equiv -C_{{\rm m}k}/C_{\Sigma k}$ entering the operator $q(n_1,n_2)$. Proceeding as outlined above we
find the effective qubit Hamiltonian
\bea
  H_{\rm eff}
  &=&\sum_{k=1,2} H_k-2E_C\bracket{0}{n-n_{\rm g}}{0}\,q
    -\sum_{m=1}^\infty\frac{2E_C^2}{\hbar\omega_{m0}^2}
    \left|\bracket{0}{n}{m}\right|^2\cdot {\rm i} [q,\dot{q}]\nonumber\\
  &&+E_C q^2\left(1-\sum_{m=1}^\infty\frac{4E_C}
    {\hbar\omega_{m0}}\left|\bracket{0}{n}{m}\right|^2\right)
    -\dot{q}^2\,\sum_{m=1}^\infty\frac{4E_C^2}{\hbar\omega_{m0}^3}
    \left|\bracket{0}{n}{m}\right|^2\ ,
\eea
where $\dot{q}=-\sum_j\frac{E_{{\rm J}j}}{\hbar}c_j \sin\phi_j$ and
${[}q,\dot{q}{]}={\rm i}\sum_j c_j^2 \frac{E_{{\rm J}j}}{\hbar}\cos\phi_j$.
By separating the single-qubit and coupling terms we arrive at $H_{\rm eff} = \sum_{k=1,2} H_k' + H_{\rm int}$.
The single-qubit terms $H_k'=H_k + \delta H_k$ acquire a small correction, $\delta H_k = O(c_k)$, while the
interaction has two contributions,
\be
  H_{\rm int} = \lambda_{\rm c}(n_{\rm g}) n_1 n_2 +
    \lambda_{\rm i}(n_{\rm g}) \sin\phi_1 \sin\phi_2\ ,
\ee
with
\bea
\label{eq:Hint_general}
  &&\lambda_{\rm c}(n_{\rm g}) = 2E_C
    \left(1-\sum_{m=1}^\infty\frac{4E_C}{\hbar\omega_{m0}}
      \left|\bracket{0}{n}{m}\right|^2
    \right)\,c_1 c_2\ ,\nonumber\\
  &&\lambda_{\rm i}(n_{\rm g})=-
    \left(\sum_{m=1}^\infty\frac{8E_C^2}{\hbar^3\omega_{m0}^3}
       \left|\bracket{0}{n}{m}\right|^2
    \right) c_1 c_2 E_{{\rm J} 1} E_{{\rm J}2} \ .
\eea
The first term ($\lambda_{\rm c}$) corresponds to a tunable capacitive coupling, while the second one
($\lambda_{\rm i}$), which arises due to the nonadiabatic corrections, corresponds to an inductive
interaction, coupling the operators of current flowing into/out of the qubit islands. This inductive coupling
is weak due to a small factor $E_{{\rm J}1} E_{{\rm J}2}/(\hbar\omega_{m0})^2$. Yet, as we show below, it
dominates, when the capacitive interaction vanishes or is small.

\section{Coupling junction in the charge regime, $E_C \gg E_{\rm J}$}
\begin{figure}
\threeimages[width=4.8cm]{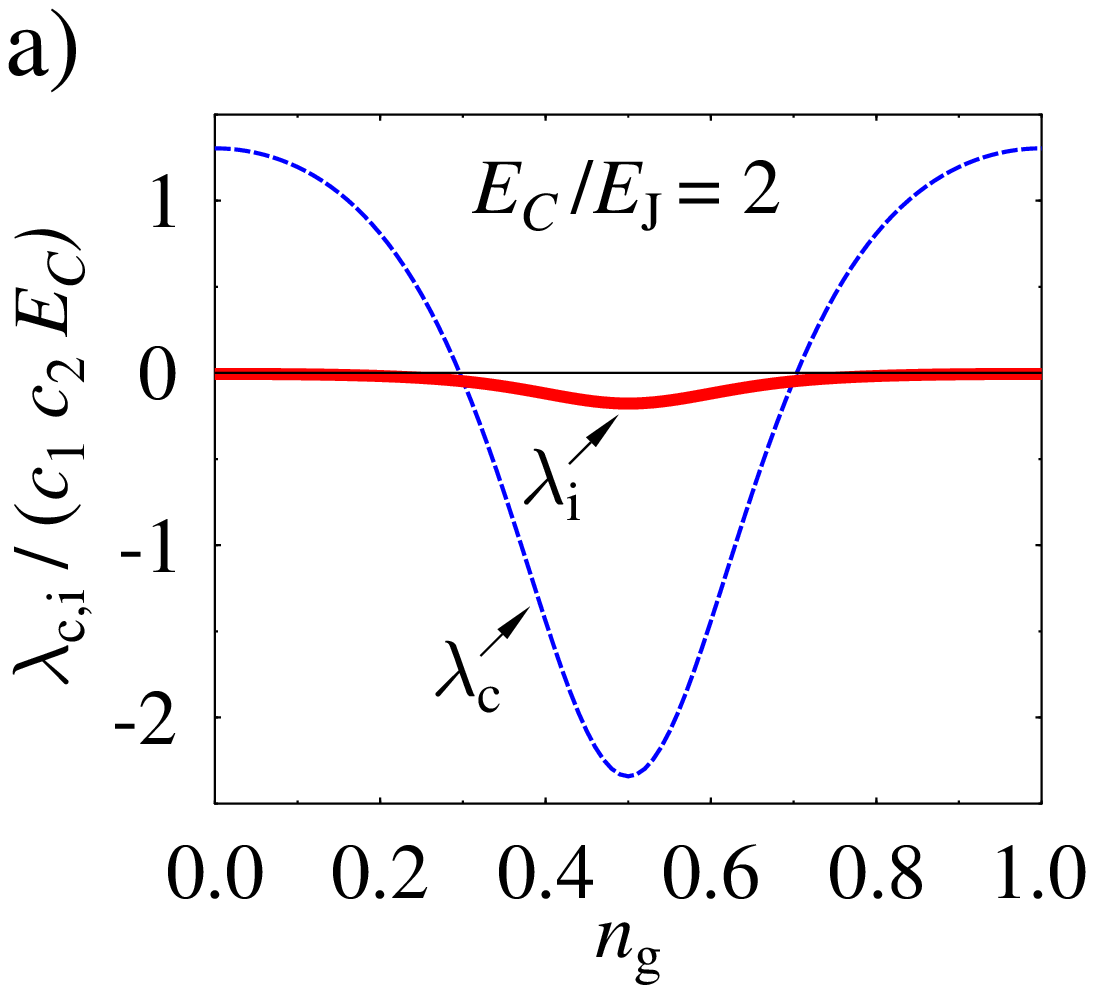}{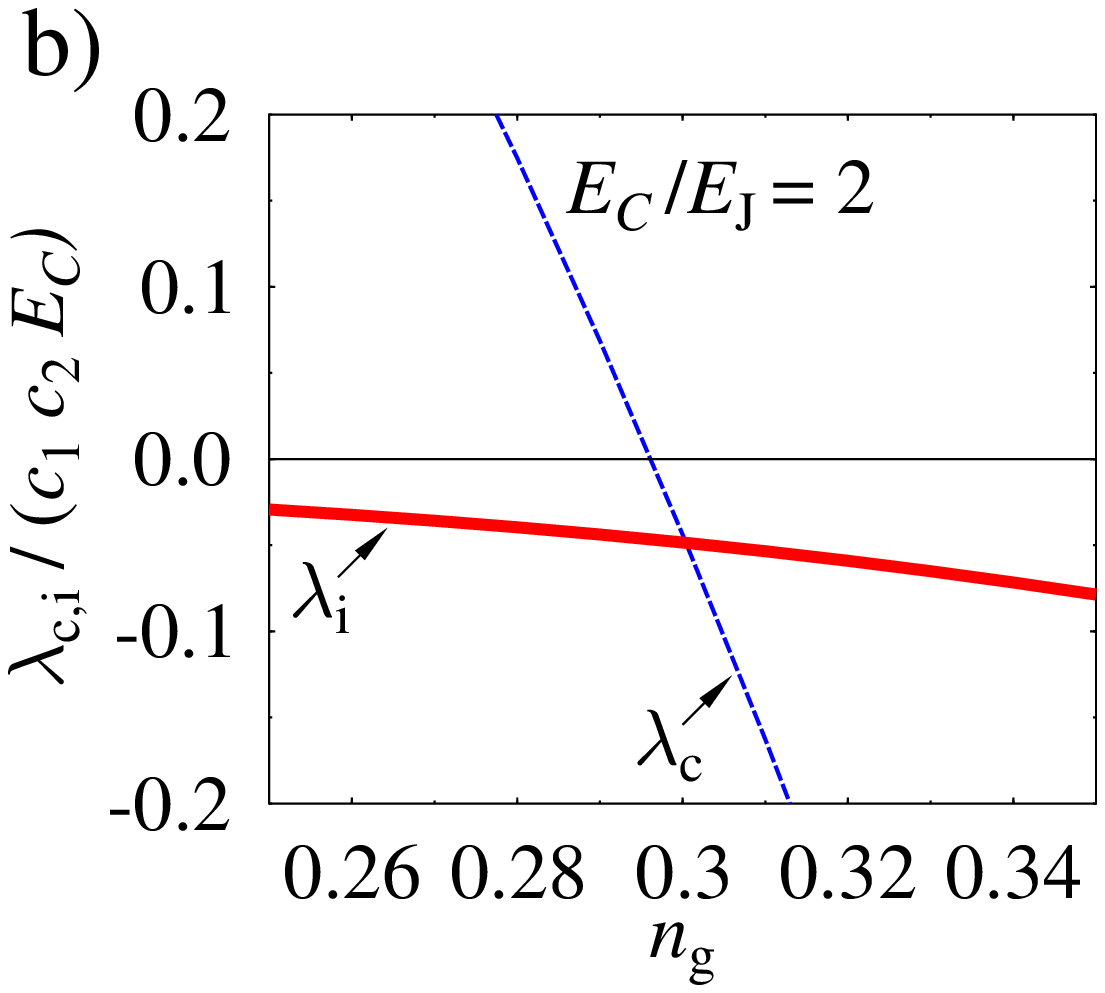}{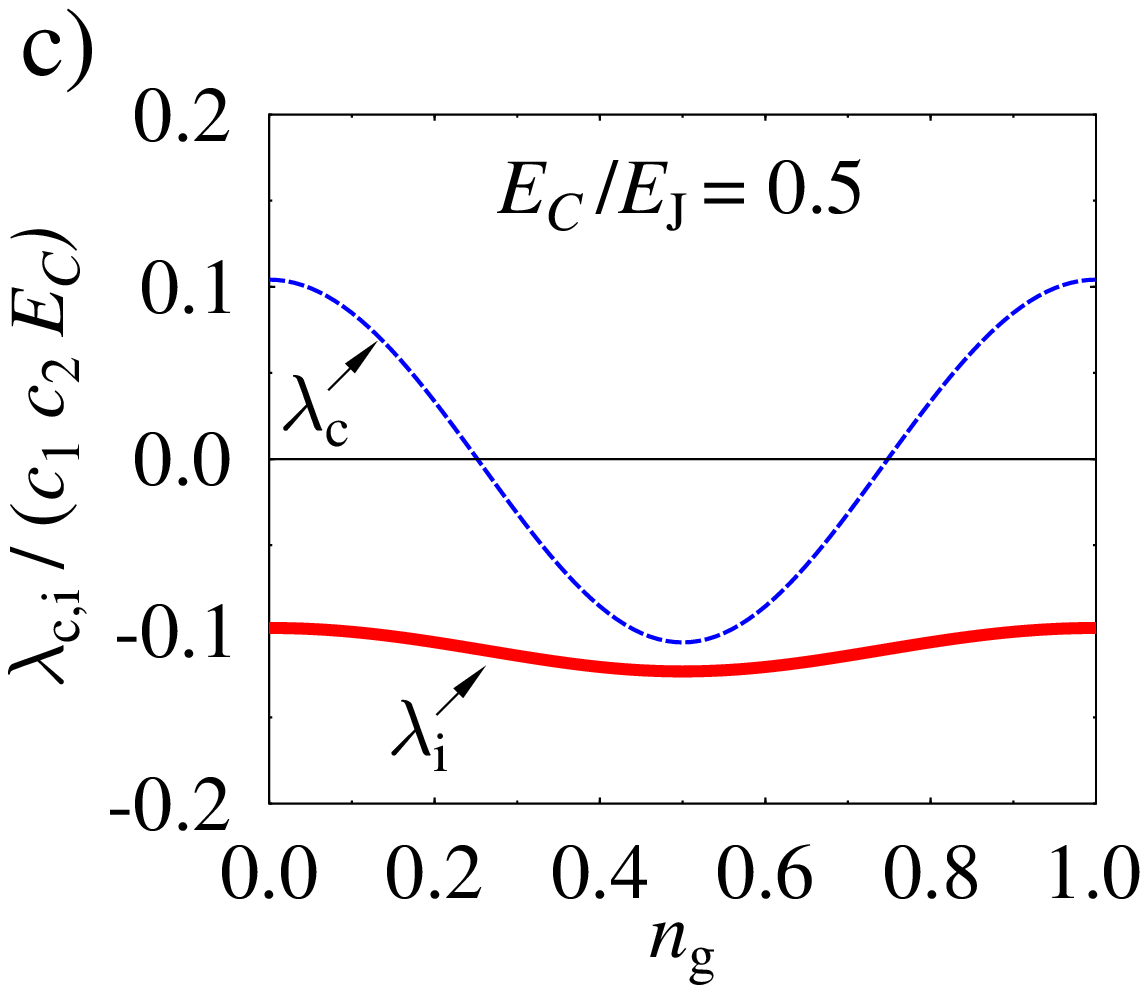}
\caption{Coupling constants $\lambda_{\rm c}(n_{\rm g})$ and
  $\lambda_{\rm i}(n_{\rm g})$ in units of $c_1 c_2 E_C$ obtained from
  eq.~(\ref{eq:Hint_general}).
  {a)} $E_C = 2 E_{\rm J}$, $E_{\rm J1}=E_{\rm J2}=0.2 E_{\rm J}$.
  {b)} A closer look at the vicinity of the ``zero" point.
  {c)} $E_C = 0.5 E_{\rm J}$, $E_{\rm J1}=E_{\rm J2}=0.2 E_{\rm J}$.
  For other values of $E_{\rm J1}$ and $E_{\rm J2}$ the inductive coupling
  constant can be obtained by simple scaling
  (see eq.~(\ref{eq:Hint_general})).}
\label{Figure:Couplings}
\end{figure}

In this limit, working close to the degeneracy point $n_{\rm g}=1/2$,
we can approximate the coupler by a two-level system with the two charge states
$\ket{n=1}$,$\ket{n=0}$ as basis. The coupler Hamiltonian then reads
\be
  H_{\rm coupler}=-\frac{1}{2}E_C(2n_{\rm g}-1)\sigma_z-
    \frac{1}{2} E_{\rm J}\sigma_x  = - \frac{1}{2}\hbar\omega_{10}\rho_z .
\ee
In the second, diagonal form we used $\hbar \omega_{10}=\sqrt{E_{\rm J}^2+E_C^2(2n_{\rm g}-1)^2} \,$,
$ \sin\eta=E_{\rm J}/(\hbar \omega_{10}) $, and $\sigma_{z}=\cos\eta\,\rho_{z}-\sin\eta\,\rho_{x}$. Expressing
the non-vanishing matrix element in eq.~(\ref{eq:Hint_general}) as $ \bracket{0}{n}{1}=-\frac{1}{2}\sin\eta$
we obtain the coupling constants
\bea
  \label{eq:Hint_TL}
  \lambda_{\rm c}(n_{\rm g})
  = 2E_C \left(1-\frac{E_C}{E_{\rm J}}\sin^3\eta\right)
    \,c_1 c_2\quad , \quad \lambda_{\rm i}(n_{\rm g})
  =-\frac{2E_C^2}{E_{\rm J}^3}\sin^5\-\eta\,\, c_1 c_2 E_{{\rm J}1} E_{{\rm J}2}
    \ .
\eea
When the capacitive interaction is switched off, at $\sin^3\eta(n_{{\rm g}0}) = E_{\rm J}/E_C$, the inductive
coupling persists
\be
  \label{eq:lambdaZero}
  \lambda_{\rm i}(n_{{\rm g}0})
  =-2c_1 c_2(E_C/E_{\rm J})^{1/3}(E_{{\rm J}1}E_{{\rm J}2}/E_{\rm J})\ .
\ee
Assuming that the ratio $E_C/E_{\rm J}$ is not extremely large, and using $E_{\rm J} > E_{{\rm J}k}$ (which
is required in the charge regime to fulfill the adiabaticity condition), we find
$
|\lambda_{\rm i}(n_{{\rm g}0})| \lesssim c_1 c_2 E_{{\rm J}k}
$.
Further away from the limit $E_C\gg E_{\rm J}$, {\it i.e.} for $E_C\gtrsim E_J$, the coupling constants
can still be calculated numerically using eq.~(\ref{eq:Hint_general}).
An example for $E_C/E_{\rm J}=2$ is shown in figs.~\ref{Figure:Couplings}a) and b).

\section{Coupling junction in the phase regime, $E_C \ll E_{\rm J}$}
In this limit the phase of the coupler $\phi$ remains small, $\langle \phi^2 \rangle \ll 2\pi$. Then we can
expand $E_{\rm J}\cos\phi\approx E_{\rm J}-E_{\rm J}\phi^2/2$, and the coupler Hamiltonian
\be
  H_{\rm coupler} \simeq E_C (n-n_{\rm g})^2+E_{\rm J} \phi^2/2\quad ,
\ee
reduces to that of a shifted harmonic oscillator. Making use of
$n-n_{\rm g}=i(a^\dagger-a)\sqrt{\hbar\omega_{10}/(4E_C)}$
with $\omega_{10}=\frac{1}{\hbar}\sqrt{2E_{\rm J}E_C}$, we evaluate $\lambda_c$ and $\lambda_i$ from
eq.~(\ref{eq:Hint_general}). It turns out, that in the considered limit and approximation the capacitive part
of the interaction vanishes while the inductive coupling constant,
$\lambda_{\rm i}(n_{\rm g}) = -c_1 c_2 (E_{{\rm J}1} E_{{\rm J}2}/E_{\rm J})$,
is independent of $n_{\rm g}$. A more precise analysis of the Hamiltonian of the coupler yields an
exponentially weak $n_{\rm g}$-dependent capacitive interaction~\cite{Averin_Bruder}, but the inductive
coupling still dominates. In this limit the setup reduces to that considered in ref.~\cite{Our_PRL}, with the
Josephson junction of the coupler playing the role of the inductance of the LC circuit with
$L=\Phi_0^2/(4\pi^2 E_{\rm J})$.
Results for the regime $E_C\lesssim E_J$ are illustrated in fig.~\ref{Figure:Couplings} c).

\section{Switching off the coupling at the symmetry point}

Eq.~(\ref{eq:Hint_general}) or eq.~(\ref{eq:Hint_TL}) show that the coupling cannot be switched off completely.
Yet, in certain situations, at least in the rotating wave approximation (RWA), the most important part of
the coupling can be switched off. As an example we consider two qubits ($k=1,2$) in the charging regime with
$  H_k=-\frac{1}{2}B_{zk}\sigma_{zk} -\frac{1}{2}B_{xk}\sigma_{xk}$.
Ignoring renormalization effects we have
$B_{zk}=E_{Ck}(2n_{{\rm g}k}-1)\sigma_{zk}$ and $B_{xk}=E_{{\rm J}k}$;
renormalization effects introduce a gate charge dependence due to voltage crosstalk,
$B_{zk}=B_{zk}(n_{g1},n_{g2},n_g)$. The effective interaction Hamiltonian can be written as
\be
  H_{\rm int} = \frac{\lambda_{\rm c}(n_{\rm g})}{4} \sigma_{z1} \sigma_{z2}
    +\frac{\lambda_{\rm i}(n_{\rm g})}{4}\sigma_{y1} \sigma_{y2} \ .
\ee
Diagonalization of $H_k$ leads to $H_k = -\frac{1}{2}\Delta E_{k}\rho_{zk}$ with
$\Delta E_{k}=\sqrt{B_{xk}^2+B_{zk}^2}$, and $\sigma_{zk}=\cos\eta_k \rho_{zk}-\sin\eta_k \rho_{xk}$ with
$\sin\eta_k = B_{xk}/\Delta E_{k}$. The interaction Hamiltonian then reads
\be
  H_{\rm int} = \frac{\lambda_{\rm c}(n_{\rm g})}{4}
  (\cos\eta_1 \rho_{z1}-\sin\eta_1 \rho_{x1})
  (\cos\eta_2 \rho_{z2}-\sin\eta_2\rho_{x2})
  + \frac{\lambda_{\rm i}(n_{\rm g})}{4}\rho_{y1} \rho_{y2} \ .
\ee
If we assume the qubits to have equal or similar energy splittings, $\Delta E_{1} \approx \Delta E_{2}$,
the most important (RWA) part of the interaction is
\be
  H_{\rm int,RWA} =
  \frac{\lambda_{\rm c}(n_{\rm g})}{4}\cos\eta_1\cos\eta_2\rho_{z1}\rho_{z2}+
  \left[\frac{\lambda_{\rm c}(n_{\rm g})}{4}\sin\eta_1\sin\eta_2 +
    \frac{\lambda_{\rm i}(n_{\rm g})}{4}\right]\,
  [\rho_{+1} \rho_{-2}+\rho_{-1}\rho_{+2}] \ .
\ee
This interaction term can be switched off completely only at the symmetry point of at least one of the
qubits, {\it i.e.}, when $\cos\eta_1=0$ or $\cos\eta_2=0 \,$, by choosing $n_{\rm g}$ such that the second
term vanishes. This means, the capacitive coupling constant has to be tuned to a value opposite in sign
but of the same order as the inductive coupling constant (instead of zero, as one might have guessed
intuitively). The cancelation is in general not possible for $E_{\rm J} \gg E_C$, when
$\lambda_{\rm c}(n_{\rm g})$ is exponentially small. We further note that at the double symmetry point
$\cos\eta_1=\cos\eta_2=0$ the decoupling is stable in linear order with respect to fluctuations of $\eta_k$.

\section{Summary and discussion}

We have analyzed the tunable coupling of qubits mediated by a fast coupling element and evaluated the
lowest-order nonadiabatic corrections beyond the Born-Oppenheimer approximation. When the adiabatic coupling
is tuned to cancel an additional fixed coupling these nonadiabatic contributions become relevant and can
even dominate.

As an example we investigated the capacitive tunable coupler proposed by Averin and Bruder, and found that
it works only in the regime $E_C > E_{\rm J}$. While the total coupling can never be switched off completely,
this is possible (in RWA) for the most important part of the coupling at the degeneracy point of both qubits.
In the opposite limit $E_{\rm J}\gg E_C$ we recovered results of an earlier coupling scheme~\cite{Our_PRL}.
Here the capacitive coupling (nearly) vanishes, while the inductive one dominates. It can not be tuned by the
coupler, unless one further modifies the design, {\it e.g.} in the way as suggested in ref.~\cite{Our_Nature}.

One strategy to keep the nonadiabatic terms weak is to design a coupler with small adiabaticity parameter
$|H_{\rm qubits}|/(E_1-E_0)$, {\it i.e.} the maximum energy difference between the eigenstates of
$H_{\rm qubits}$ should be much smaller than the minimum excitation energy of the coupler.
Another strategy is, to develop designs where $[B,H_{\rm qubits}]=0$ and the Born-Oppenheimer approximation
becomes exact and, hence, no nonadiabatic corrections appear. Such designs have been
proposed~\cite{You_commuting,Lantz}, with tunability achieved via the control of circulating currents in
SQUID loops. However, the coupling can be switched off only if the system is fabricated with identical
Josephson junctions in the SQUID loops of the qubits, which is difficult to realize experimentally.

\acknowledgments

We thank A.~Zorin for fruitful discussions.
This work is supported by the Landesstiftung Baden-W\"urttemberg gGmbH and
further by the EU IST Project EuroSQIP, as well as by the Russian Science Support
Foundation (YM). CH was supported by Graduiertenkolleg ``Kollektive Ph{\"a}nomene im Festk{\"o}rper".

\bibliographystyle{mybstEPLshort}
\bibliography{ref}

\end{document}